# Barriers to device longevity and reuse: A vintage device empirical study

Craig Goodwin [0000-0003-3296-7200] and Sandra Woolley [0000-0002-7623-2866]

Keele University, Staffordshire, UK
c.goodwin@keele.ac.uk

**Highlights:**

- Barriers to app installation impede device longevity.
- Relatively few apps could be downloaded directly onto a vintage device.
- Many apps could be downloaded indirectly.
- 61.3% of apps (downloaded directly or indirectly) were capable of functioning.

**Abstract.** This extended paper contributes a methodology and a detailed analysis of app installation and functionality on a 'vintage' device. Experimental results are presented that demonstrate barriers to the reuse of vintage Apple devices, and solutions are posited. 230 apps across 23 unique app categories were tested to determine if they could be downloaded, installed, and opened, and whether they appeared functional on a vintage Apple device. Only 29 (12.6%) of the apps could be downloaded directly, and in contrast 140 (60.9%) of the apps were downloadable with the aid of another Apple device. In total, 141 (61.3%) of applications downloaded either directly or indirectly were considered functional and capable of running on the device. We discuss measures Apple and developers could take to support legacy devices users, prolong device use, enable reuse and, potentially, prevent functional devices from becoming e-waste.

## 1 Introduction

Approximately, 53 million metric tonnes of electronic waste are produced every year, and this is expected to rise to 75 million tonnes by 2030 [1]. Despite many electronic devices being recyclable, only 15-20% are typically recycled and most are disposed of in landfill after being discarded because of obstacles to their continued use [2-3]. Barriers to repair include limited device warranties and the need for specialist technical skills or tools to repair devices.

Apple Inc. is the market leader in the technology sector with an estimated market capitalization of 3 trillion US dollars [4]. New models of Apple iPhone smartphones have been released every year since 2007 [5] and Apple's tablet devices have been updated on an annual basis since 2010 [5]. Schemes such as the iPhone Upgrade Plan [6] enable consumers to receive new iPhones on a yearly basis and return their 'old' device [7]. Apple publish annual Environmental Progress Reports and have announced a commitment to a 2030 carbon neutral goal [7, 8] but little is known in general about the second life of refurbished devices, and of devices that are recycled or disposed of in landfill.

The motivation for the study presented here was to explore device obsolescence and glean insights into the extent of challenges of continued device reuse. The reuse of a device in this context is defined as the use of a device that is beyond its initial usage phase, either via the changing of ownership or following significant time in dormant storage [40, 41]. Significant barriers to continued reuse of older devices are i) a lack of compatibility with new and updated applications and ii) a lack of information about compatibility. App store application details about compatibility whilst generally useful, aren't always a reliable source of information for vintage devices users. Importantly, at the time of writing, the only method of discovering whether an Apple iOS application functions on a vintage device is to manually attempt downloading each application. The study presented in this paper exemplifies the challenge of continued device use with a 'vintage' Apple Mini tablet. This paper expands upon a previously published conference paper [38] by providing additional background on related works, additional figures supporting the study method and results, and an expanded discussion.

## 2 Background

There is a lack of research and innovation in legacy device use and reuse. Devices can be defined as legacy if they are, for example, an end-of-life product, no longer supported by the supplier or updates are no longer available [39]. There is little academic literature on end-of-life devices, but in contrast there is a significant amount of grey literature discussing device reuse which helps contextualize the research area. For example, there are consumer guides for repurposing and reusing older smart devices, for example, using devices as notepads or for cooking instructions [10-14], but these don't generally involve meaningful reuse strategies (i.e., actual device use/reuse or ethical recycling). Another example is that of device ownership. There are many reasons for device ownership longevity, but iPhone users are more likely to sell or trade their old device than Android users [15]. Approximately 30% of iPhone users keep their old devices irrespective of lifecycle age [15], which is a significant portion of the consumer base. Statistics such as these highlight the importance of researching the longevity and use/reuse of older devices [21] (and furthermore topics such as functionality, accessibility, and e-waste) in the wider context of the circular economy [16].

Li et al (2010) [17] addressed the "dark side of Moore's Law" in the first work in academic literature to address smartphone evolution and reuse and posit a more sustainable model. More recently there has been much more discussion regarding device reuse and reuse models [17]. Of course, this can be due to the reason that smart devices are only now starting to become "obsolete" in use and therefore strategies to continued reuse or sustainable alternatives need to be made [18, 19]. Boano [20] proposes increasing the durability and support of legacy devices to "avoid an 'Internet of Trash'" concluding that it is *"important to study how to increase the lifetime of IoT devices that have become obsolete or no longer compatible with the latest communication standards, so to prevent an early disposal."* Safe disposal of these devices is also a contentious issue, as often sensitive data can be retrieved from devices sold by consumers [22], sent for refurbishment [23] or even recycled [21, 24, 25].

Despite recent interest in app store business models, for example, as evidenced by the proposed US Congress Open Act Markets antitrust bill and the Fortnite vs. Google lawsuit [26], there is a persistent lack of research and incentivization for manufacturers and developers to increase device lifespans and extend device support. This furthers the need for research exploration in the practice and drivers of 'sideloading', which is defined as the download and installation of applications outside of official smartphone application marketplaces [27].

### 2.1 Legacy Device Support

There is little consensus on the best approaches to device reuse of vintage smartphones, tablets, and smart devices. The current accepted model involves the sustainable recycling of devices, but this solution doesn't work for those who want continued reuse of their devices. Once manufacturer support has ended, it is often down to niche internet communities to continue third-party support for legacy devices. However, it can be more difficult to increase longevity for devices with closed ecosystems (such as Apple) and in the absence of official documentation supporting vintage devices, barriers to continued use become apparent. Because of this, vintage device use irrespective of device age may resort to third party (and often unreliable or outdated) advice.

When going to the Apple App store with a current/vintage/obsolete Apple device, all the applications are available to download, therefore the user must manually figure out which applications are compatible. In comparison, the applications on the Google Play Store target supported devices with supported manifests and device features [31]. This dynamic method of delivering applications, referred to as 'conditional delivery" [32], allows the user to know what applications are supported (and ones that aren't). There are exceptions to this, such as apps that have manifests that aren't correctly programmed for specific devices, or simply applications that (although they do support the device) are removed from the Google Play store for specific devices.

There are three solutions to downloading previously supported applications on older Apple devices (see Table 1). The first is being fortunate enough to have a pre-existing Apple account with purchase (or download) history for that application [33]. The second is to use an additional device with a modern version of iOS to download the required application then use the same account with the purchase history on the older device to directly download the older last supported version of the application [34]. Figure 1 shows an example of the steps required to download a "non-supported" application. Lastly, consumers can download via a PC or Mac an old version of Apple's "iTunes" and access the last previously supported application that way. Nevertheless, each solution either requires a history with the application or another device (or both). Sideloading applications on Apple devices is a difficult procedure for those without prior knowledge of safe sideloading and there are much fewer third-party repositories for official Apple Device applications in comparison to Google Android [35].

Further alternative approaches do exist but are not unreliable. For example, there are websites that provide direct download app links to the App Store for the last supported versions for given devices; however, these websites are generally user-managed websites and not official or reliable solutions. This means that the only official and safe working solution is via official Apple avenues.

**Table 1** Summary of the three download methods

| Methods | Summary |
|---|---|
| Direct Download | If an application still supports the targeted device, then the user can directly download the application with no issues. |
| Download via Another Device | If an application cannot be downloaded directly, then the use of pre-existing purchase history or the use of another device is required. This then allows the download of the last previously supported version of the targeted application (if the targeted app version still exists. |
| Direct Download Link | Some independent/community websites contain direct app download links for previously supported applications. This method removes the need for a prior purchase history and another non-vintage device. This app download method is uncommon and unreliable in that many websites no longer exist or are not well supported. |

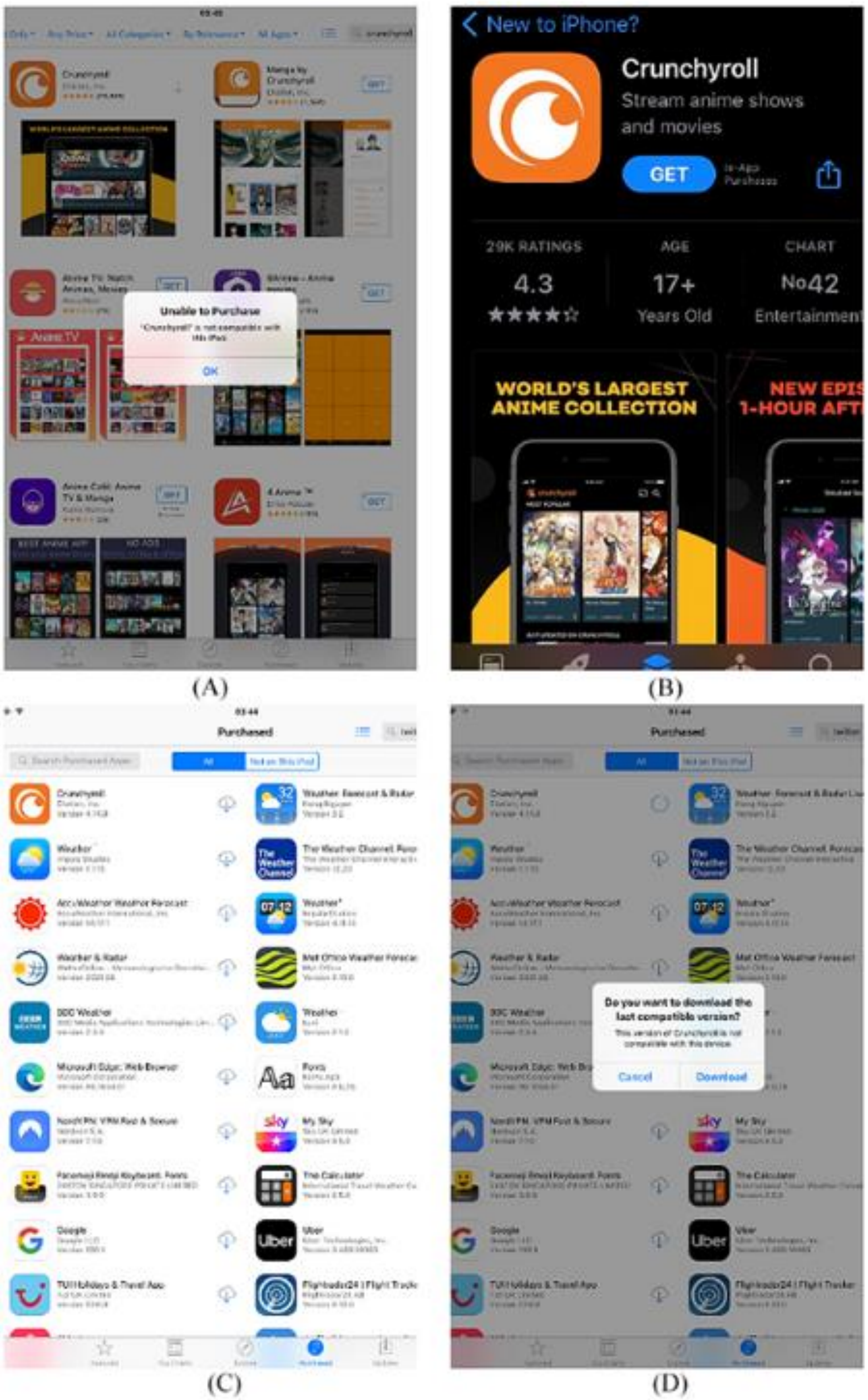

**Fig. 1**. Vintage device screenshots illustrating hurdles to downloading an application via another (non-legacy) device. (**A**): Unable to download the application directly on the iPad Mini. (**B**): Search and download the application on a supported modern Apple device with the same account credentials. (**C**): On the iPad Mini, search for the application purchase history. (**D**): Download the last compatible version of the required application (if supported).

# 3 Methods

This study aimed to investigate the barriers to device and software usage and the obstacles to app installation that users of vintage devices may experience.

**The vintage device:** The vintage device used in the study was an Apple iPad Mini tablet (iOS version 9.3.6) approximately nine years old at the time of the study which took place September 30th to October 2nd, 2021. The study was performed with a 2012 iPad Mini which is significant due to it being the last 32-bit Apple Tablet [5]. This era (2012) positions the device as "vintage" but not "obsolete" as Apple state that "Products are considered vintage when Apple stopped distributing them for sale more than 5 and less than 7 years ago" [36,37].

**The non-vintage device:** For indirect downloading, an Apple iPhone SE (iOS version 15.1, 'currently supported' at the time of the study) was used. This device allowed us to download the required application (sharing the same account with the Apple Mini) and then use the Apple Mini to view the account purchase history to download the last supported version of that application (if supported and applicable). From there, we can check whether the application successfully downloads, installs, and functions on the iPad.

## 3.1 Application Selection Criteria and Collection

Popular free application categories (23 in total) from the Apple App Store were selected. The application category "Kids" was excluded because the category consolidates other app types into one category. In addition, categories which are now included on modern Apple devices ("Apple Watch Apps", "AR Apps", "Developer Tools", "Graphics & Design" and "Safari Extensions") were excluded because they did not feature on the Apple Mini Tablet App store.

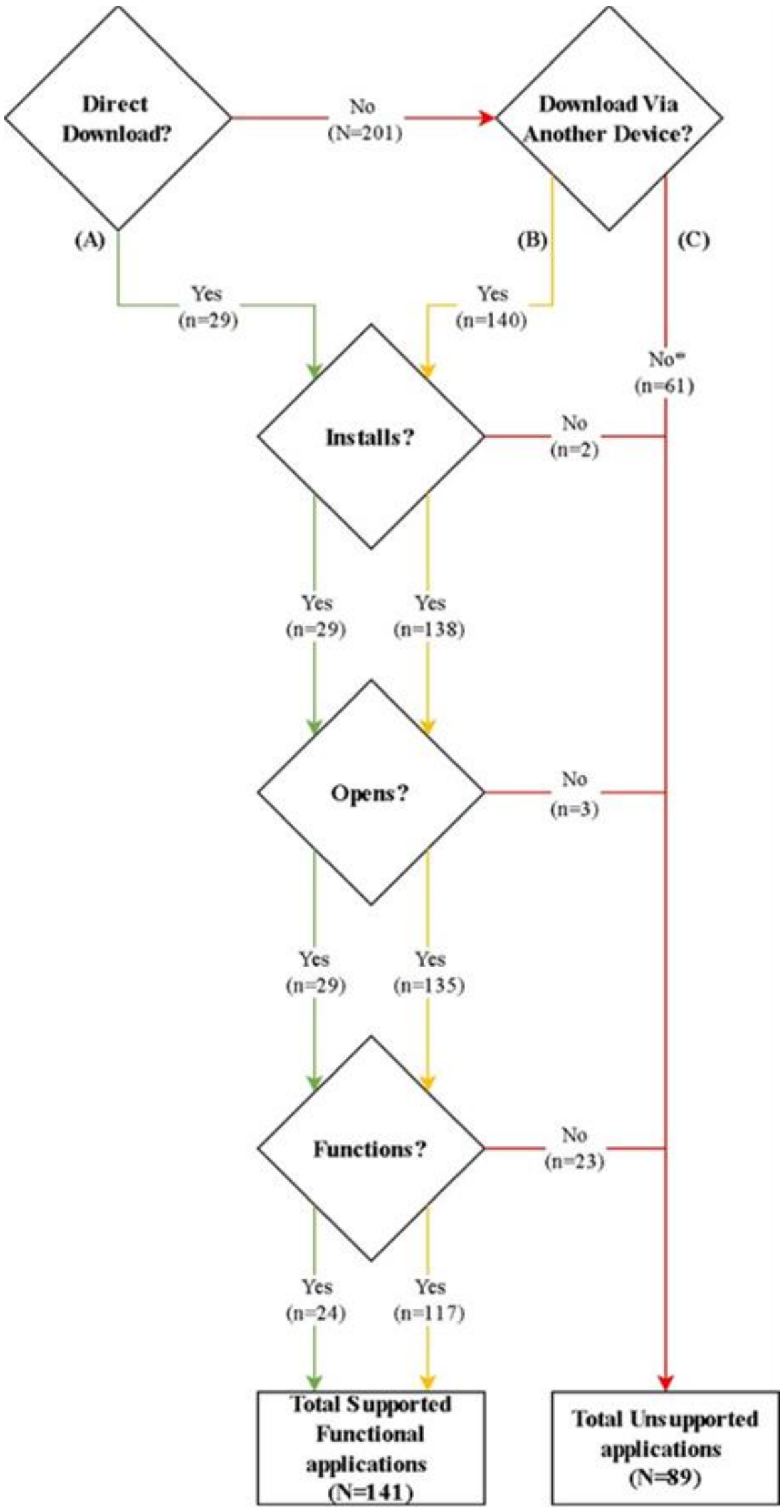

**Fig. 2.** Flowchart outlining the study method and results.

(A): Applications downloaded directly (green). (B): Applications downloaded via another device (yellow). (C): Applications that could not be downloaded directly via another device (or no previous support for the test device (red) and that failed “Installs?”, “Opens?” and “Functions?” (*34 of the applications were never supported by the device iOS version.)

| **BOOKS:** | **GAMES:** | **NAVIGATION:** | **SOCIAL NETWORKING:** |
|---|---|---|---|
| Kindle | Truck Simulator | Google Maps | Messengers for WhatsApp |
| Audible | Pokémon Unite | Street View Map 360 | Facebook8 |
| Teach Monster | The Ants: Kingdom | what3words | Messenger |
| Adobe Digital Editions | Get Lucky 3D | SpTH | Messenger for WhatsApp |
| Wattpad | Roblox | Waze | Discord |
| Libby | Dodge Agent | iStreets | Skype |
| BorrowBox Library | Hand Strike | OS Maps | Hangouts |
| Google Play Books | Make 'Em | Tube Map | Whats up for WhatsApp |
| iReader | Switch Race! | TfL Go | Google Duo |
| VleBooks ebook Reader | Rope-Man Run | Moovit | Telegram |

| **BUSINESS:** | **HEALTH & FITNESS:** | **NEWS:** | **SPORTS:** |
|---|---|---|---|
| Microsoft Teams | Airmid UK | Twitter | BT Sport |
| Zoom Cloud Meetings | PhysiApp | BBC News | Sky Sports |
| Adobe Acrobat Reader | Headspace: Meditation | Reddit | FUT 22 Draft |
| Google Meet | Les Mills+ | Nextdoor | BBC Sport |
| Cisco WebEx Meetings | Calm - Medication | Sky News | LiveScore |
| Indeed Job Search | MyFitnessPal - Calorie | Mail Online | Premier League |
| LinkedIn | TeasEar | NewsBreak | DAZN |
| Adobe Scan | Calorie Counter+ | Financial Times | The FA Player |
| GoToWebinar | FitOn Workouts | PressReader | Pacwyn 21 |
| Officesuit & PDF editor | White Noise Deep Sleep | The Economist | FUTWIZ FUT 22 |

| **EDUCATION:** | **LIFESTYLE:** | **PHOTO & VIDEO:** | **TRAVEL:** |
|---|---|---|---|
| Times Tables Rock Stars | Clock - Clock App | YouTube | Google Earth |
| Google Classroom | Pinterest | Adobe Photoshop Express | Trainline |
| Toca Life World | Amazon Alexa | Capcut | Booking.com |
| Duolingo - Language | Google Home | Picsart | Jet2Holidays |
| World of Peppa Pig | Super Slime Simulator | Twitch | Airbnb |
| Blackboard | Rightmove Search | Canva | Flightradar24 |
| Sumdog | The National Lottery | Google Photos | TUI Holidays |
| Seesaw Class | VoiceGaga | PicCollage | Jet2.com |
| Toca Hair Salon 4 | Sendit | FreePrints | British Airways |
| CodeSpark Academy Kids | Zinnia | Canon Printjet | Uber |

| **ENTERTAINMENT:** | **MAGAZINE & NEWSPAPER:** | **PRODUCTIVITY:** | **UTILITIES:** |
|---|---|---|---|
| Netflix | The Guardian | Microsoft Word | Google Chrome |
| Disney+ | The Telegraph UK* | Microsoft Outlook | Google |
| All 4 | Daily Mail | Microsoft PowerPoint | The Calculator |
| Amazon Prime Video | Daily Mirror | Gmail | Calculator Air |
| BBC Iplayer | The Times * | Microsoft Office | Facemoji |
| TikTok | Readly | Microsoft Excel | My Sky |
| Sky Go | The Guardian Editors | Microsoft OneNote | Calculator |
| ITV Hub | Daily Record Newspaper | Google Docs | NordVPN |
| YouTube Kids | The Independent | Microsoft OneDrive | Fonts |
| Paw Patrol Rescue World | Speechify | Google Drive | Microsoft Edge |
| **FINANCE:** | **MEDICAL:** | **REFERENCE:** | **WEATHER:** |
| Barclays | NHS App | Google Translate | Weather (luni) |
| Virgin Money | Well Repeat | Night Sky | BBC Weather |
| Lloyds Bank | Patient Access | Oxford Dictionary English | Met Office |
| Santander | LloydsDirect | Translate Now | Weather & Radar |
| NatWest | BNF Publications | Dialog | Weather+ |
| Halifax | Covid-19 and flu info | Daily Random Facts | Accuweather |
| Nationwide | 2021 Atlas Perpetual | Bible | Weather Free |
| HSBC | SystmOnline | Mods & S for Among Us | The Weather Channel |
| Go Henry | TeachME Anatomy 3D | Addons for Minecraft PE | Weather (Weather) |
| Starling Bank | Livi - See a GP Video | Crafty Craft for Minecraft | Weather Forecast |
| **FOOD & DRINK:** | **MUSIC:** | **SHOPPING:** | |
| Uber Eats | Spotify | Amazon | |
| Deliveroo | BBC Sounds | Shein | |
| Just Eat | Amazon Music | Asos | |
| HelloFresh | Beat Maker Pro | eBay | |
| Domino's Pizza | YouTube Music | Vinted | |
| Morrisons Groceries | Piano - Play | Tesco Grocery | |
| Too Good to Go | Guitar - Real Games | M&S | |
| Costa Coffee Club | Drum Pad Machine | Etsy | |
| BBC Good Food | Shazam | Argos | |
| Gousto | The Piano Keyboard | BooHoo | |

**Fig 3:** The 230 applications for each of the 23 selected major categories in the Apple App Store *(*These applications also featured in the 'News' Category, so the next original application was chosen.)*

After selecting the 230 applications for download, the following protocols were prepared for each application to follow in order. For example, if an application directly downloads, installs but doesn't open, we can assume the application does not function and therefore does not meet the criteria for a "functioning supported application". There are many different interpretations of functionality and "functional applications" which are more thorough and systematic in measurement (such as JIT mechanisms). For this study, functionality was difficult to measure without extensive resources therefore the most basic measurement method was applied:

- **Direct Download:** Can the application be downloaded directly via the App Store (regardless of the application manifest requirements?)
- **Download via another device:** Regardless of whether the application can be installed directly or not, can the application be installed to another modern device to then access the purchase history on the older device to download the last supported application for that device?
- **Installs:** Does the application install correctly?
- **Opens:** Does the application launch without force quitting or similar issues?

- **Functions:** Does the application function as intended? (For practical purposes some applications were assumed to function if they open without closing. For example, a banking application would be assumed to function if it opens without closing, but for obvious reasons you cannot login or test functionality beyond this point without credentials.) This step assumes the most minimum of usability testing protocols without extensive application testing.
- **Previous App Support:** Has the application ever been supported by the device in the past (regardless of whether it isn't supported currently)?

# 4 Results

As shown in Figure 4 and a summary of the results in Table 2, the study shows a large and distinct difference between applications that can be downloaded directly vs applications that can be downloaded via the help of another device vs applications that couldn't be downloaded either way (**~12.6% vs ~60.9% vs ~26.5%**). There was also a high uptake of previous application support, as only 31 applications were never compatible with the study device. As shown in Figure 4, the categories with the highest number of applications that could be directly downloaded were "Magazines" and "Books" (**5 and 4**) which is to be expected as these are less device intensive applications and therefore can usually be supported by developers for a longer period.

**Table 2:** Summary of results

| | **Percentage of Applications** |
|---|---|
| Directly Downloadable | 12.6% |
| Downloadable Via Another Device | 60.9% |
| Neither | 26.5% |
| Directly Downloadable and Functions | 10.4% |
| Downloadable via Another Device and Functions | 50.9% |
| **Total Supported Functional Apps** | **61.3%** |
| **Total Unsupported (Unfunctional) Apps** | **38.7%** |

Nine of the categories had no applications which could be directly downloaded, and likewise a further seven categories only had one application available to download.

However, there is a huge increase in certain categories for applications that can be downloaded with the help of another device, therefore enabling the Apple Mini to download the last previously supported version of the application. The 'Productivity' category had zero applications which could be downloaded directly; however, all ten applications were able to download via the help of another device. Likewise, the 'Music', 'News' and 'Photo & Video' had a similar high download success when using another device. All but three categories had a **50**% or above download success.

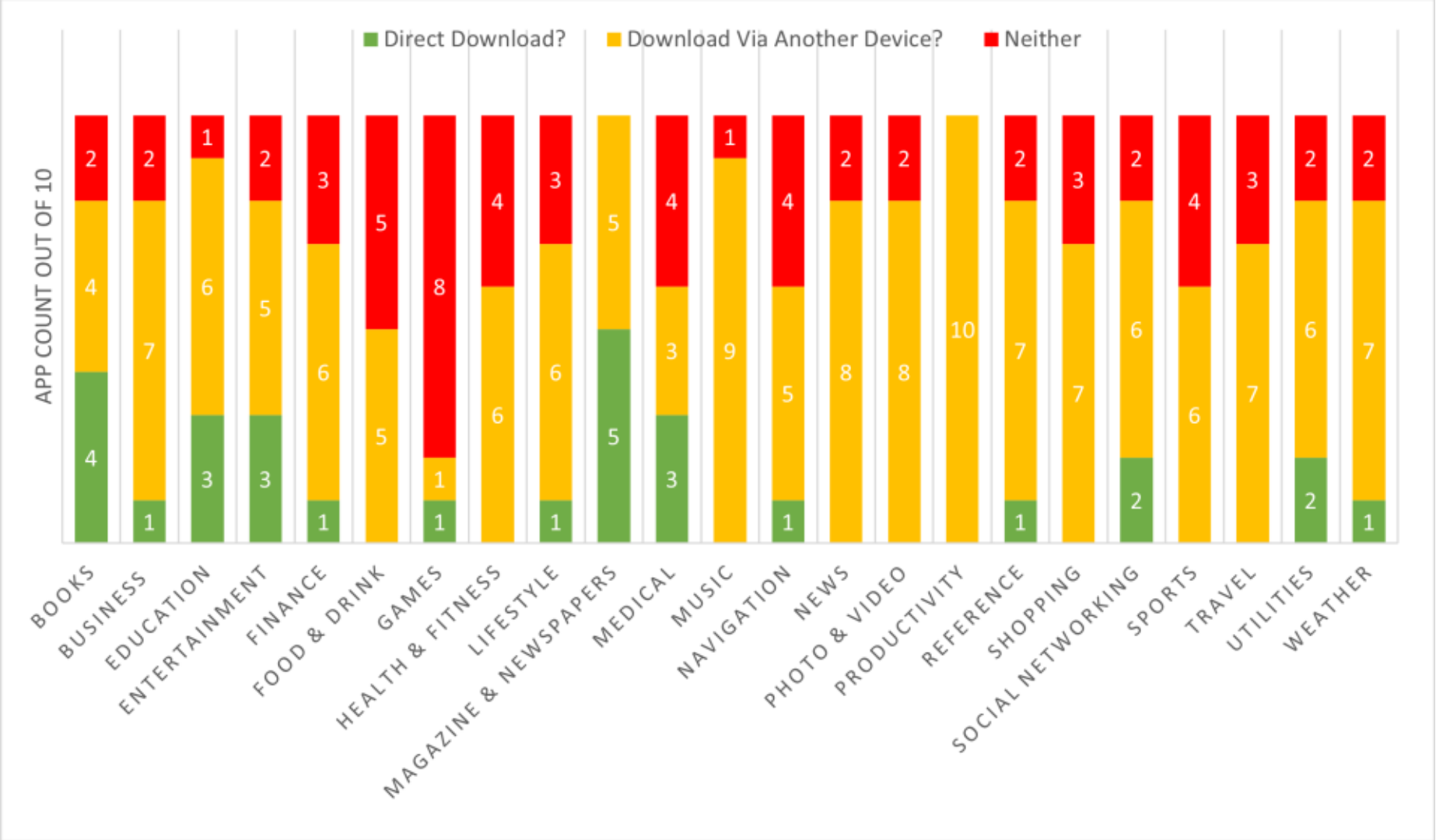

**Fig. 4.** Number of top 10 applications downloadable directly, applications downloadable via another device and applications that were not downloadable by either method.

### 4.1 Functional Supported Applications and Unsupported Applications

In total, 141 functional supported applications could be installed in comparison to the 89 unsupported applications (**61.3% vs 38.7%).** These figures indicate a much lower percentage compared to applications that could be downloaded but are still indicative of a device that has a good level of application support even if it is defined as "vintage". In order of highest functionality (see Figure 5), the application categories of (Music, News, Business, Photo & Video, Travel, Finance, Health & Lifestyle, Shopping, Sports, Food & Drink) had no applications that would function when directly downloaded, but a larger uptake in application functionality when downloaded via another device. In addition to this, the application categories of (Reference, Weather, Education, Magazines & Newspapers and Books) had at least **80**% functionality when combining those applications which can be downloaded directly and those with another device. The application categories with the lowest count were "Food & Drink" and "Games" with **10**% of applications functioning.

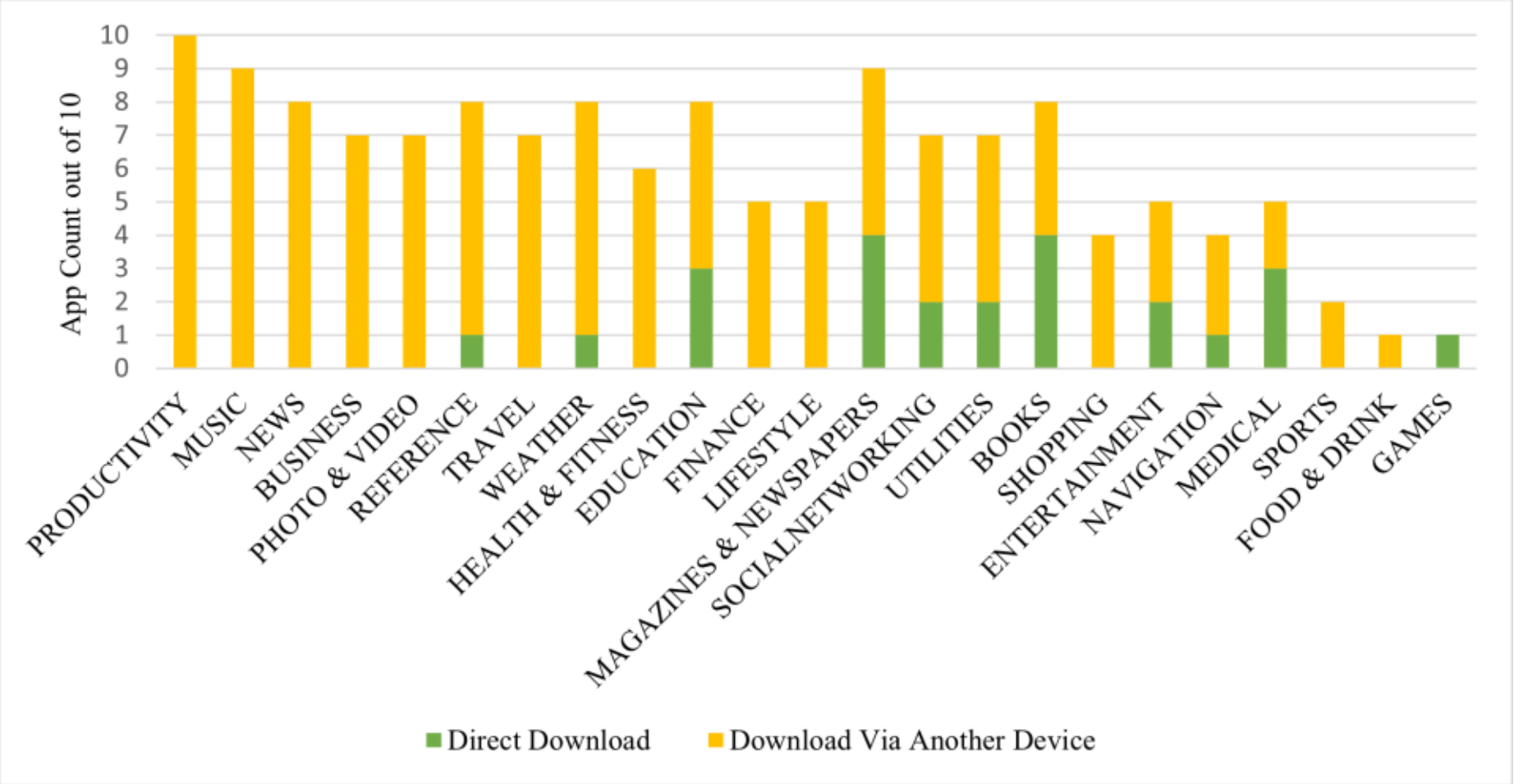

**Fig. 5.** Supported functional applications sorted by categories and stacked by download method.

# 5 Discussion

The findings of the study demonstrate that vintage Apple devices have substantial functional capability and potential for extended use. However, the majority of functional applications required the assistance of another device to enable application download. Furthermore, many of applications stated that they no longer supported the study device, but this study demonstrates that many of the previously supported versions are still fully functional. The hurdles that are implemented into the App Store and the lack of official documentation to help consumers prolong device usage is something that either Apple or application developers could address as more devices become 'vintage'.

Some downloadable applications, such as the BBC iPlayer application, display a *"device version no longer supported"* message on launch followed by a recommendation to *"watch iPlayer via a browser".* At the very least messages of this type are informative and alternative recommendations support users in finding solutions to achieve their tasks. Workarounds such as using a browser instead of native applications provide temporary solutions to problems with application support but as with most applications, even browsers will become non-functional in the future for older devices. Furthermore, the findings showed that some applications constantly reminded the user to update the application but on discovery found that the application cannot update as it is not supported, yet the application remind-ed remains.

Finally, although the main purpose of the research was to observe the difference between applications that can be download directly vs. downloaded via the help of another device, the main goal was to shine a light on barriers to continued device usage and the opportunities potential for extended-life Apple devices in the future. In the next few months, the vintage device used in the study becomes "obsolete" according to Apple's definition [36] and, one might expect, inevitably lose more application compatibility.

### 5.1 Limitations and Future Work

Due to the limited number of applications selected (only the ten most popular apps per category), the results highlight the differences in downloadable and functional applications for a relatively small sample of all applications. The difficulty in increasing the sample size for the study is the timeframe required to perform the necessary tasks for each app and the risk of applications being updated and removed from the App Store. Another difficulty was defining and determining functionality for some applications, especially within categories such as Finance and Medical where, without pre-existing accounts or records, functionality was necessarily determined against a minimum set of defined criteria.

Future work will aim to replicate the study to observe changes over time in application downloadability and functionality for the same device. It would also be interesting to perform the study with older (but still vintage) versions of different 32-bit and/or 64-bit iOS devices. Additionally, a similar methodology could be used for a study with older Android devices with sideloading as the alternative to using another device. Replicating a similar study on a similar era Android device and then comparing the results with this study may allow the discovery of differing device and application obsolescence rates. This could provide insights about the relationships between the type of devices, their ecosystems (closed vs open) and their longevity.

## 6 Conclusion

The study highlighted the barriers in place to the extended use and functionality of vintage Apple devices. In downloading and inspecting the functionality of 230 free applications across 23 application categories, this study demonstrates a substantial difference between the numbers of applications that could be directly download (29) against the applications that could only be downloaded via the help of another device (140). This difference not only demonstrates issues within Apple's App store, but also highlights the need for developers to either remove previously supported apps or continue to support vintage devices.

Whilst the majority of iPhone and iPad users will upgrade to newer hardware iterations of Apple products, there will still be a sizable contingency who remain users of older devices (using iOS 12 or earlier). It is recommended that application developers and Apple Inc. increase the transparency of application compatibility for vintage device users. It is also recommended that App store access to incompatible or non-supported applications on these devices is closed.

In conclusion, this study does not attempt to measure the functionality of each individual apps nor to measure the functionality of different application categories. Instead, it highlights the issues that legacy Apple device users face when attempting to use or reuse their device. This problem can only be resolved if measures are implemented to remove these unnecessary barriers. Research and testing of vintage and obsolete devices using this methodology is recommended to further explore the use and reuse of devices and software to underpin continued device use/reuse.

## Appendix

### Additional App Selection Criteria Detail

• Apps were required to have cross-platform support, i.e., apps exclusive to a subset of Apple devices were not selected.
• The app appeared as 'available to download' on the iPad Mini App Store.
• The app was in the Top 10 free apps of its respective category (e.g., Travel.) as of September 30th to October 2nd, 2021 (these were identified on September 30th as Apple
change the Top 10 list regularly).
• No attention made to specific application or iOS requirements. Current
or previous app support will be defined and considered after the study is completed.
• The app categories "Magazines & Newspapers" and "News" often contain the same apps. The top 10 "Magazines & Newspapers" apps were selected first and
then the top 10 of "News". Where a duplicate existed then the next available app was selected.

**Authorship Contribution Statement:**

**Craig Goodwin:** Conceptualization, Methodology, Investigation, Writing – original draft Writing – review and editing, Data Analysis
**Sandra Woolley**: Research supervision, Writing – review and editing, Data Visualization.

**Funding Sources:**
This research did not receive any specific grant from funding agencies in the public, commercial, or not-for-profit sectors.

**Vitae:**
**Craig Goodwin** is a PhD researcher at Keele University in the School of Computer Science and Mathematics. He received his bachelor's degree from the University of St Mark and St John, Plymouth, and his master's degree from London Metropolitan University. His research focuses on sideloading on smartphones and the associated drivers, motivations, and behaviors.

**Sandra Woolley** is a Deputy Director of the Digital Society Institute at Keele University and a Reader in the School of Computer Science and Mathematics where she also leads Software and Systems Engineering Research. Her research interests encompass aspects of software engineering, human-computer interaction, and health and cultural informatics.